\documentclass[RAM,USenglish,11pt]{article}

\usepackage[pdftex]{graphicx} 
\usepackage{setspace}
\usepackage[utf8]{inputenc}
\usepackage{sectsty}
\usepackage[compact]{titlesec}
\usepackage[margin=1.1in]{geometry} 
\usepackage{enumitem}
\usepackage[margin=11pt, font={sf, small}, labelfont={sf,bf}]{caption} 
\usepackage{color}
\usepackage[sf, center]{subfigure}
\usepackage{tabularx,colortbl} 
\usepackage{changepage}   
\usepackage{framed}
\usepackage{xcolor}
\usepackage[hidelinks, colorlinks=true, linkcolor=gray, citecolor=gray, urlcolor=gray]{hyperref}
\usepackage{amsmath}
\usepackage[backend=bibtex,hyperref=auto,style=numeric,defernumbers=true,maxnames=99,doi=true,isbn=false,url=false]{biblatex}
\usepackage[vskip=0pt,font=itshape]{quoting}

\bibliography{informationcollage.bib}
\bibliography{ic.bib}

\allsectionsfont{\normalfont\sffamily\bfseries}

\titlespacing{\section}{0pt}{*3}{*0}
\titlespacing{\subsection}{0pt}{*3}{*0}
\titlespacing{\subsubsection}{0pt}{*2}{*0}

\setlist{noitemsep} 
\setitemize[0]{leftmargin=18pt,itemindent=0pt}

\def\cf{\emph{cf.}}

\def\etal{\emph{et al.}}

\usepackage{datetime}
\newdateformat{monthyeardate}{%
  \monthname[\THEMONTH] \THEYEAR}

\newenvironment{techReport}
{\thispagestyle{empty}%
\begin{center}
{\LARGE\textbf{Collecting and Structuring Information in the Information Collage}}\\[.3cm]
{\large\monthyeardate\today}\\[1em]
Sebastian Sippl$^{1,2}$, Michael Sedlmair$^3$, Manuela Waldner$^1$\\ [1em] $^1$Institute of Visual Computing and Human-Centered Technology, TU Wien\\$^2$Center for Digital Safety and Security, Austrian Institute of Technology\\ $^3$Institut für Visualisierung und Interaktive Systeme, Universität Stuttgart\\ [2cm]
\end{center}
}

\begin{document}

\begin{center}\huge
  Technical Report
\end{center}

\begin{techReport}

\newenvironment{Abstract}
{ \section*{Abstract} \addcontentsline{toc}{section}{Abstract} \begin{spacing}{1.2} }
{ \end{spacing}  }

\begin{Abstract}
Knowledge workers, such as scientists, journalists, or consultants, adaptively seek, gather, and consume information. These processes are often inefficient as existing user interfaces provide limited possibilities to combine information from various sources and different formats into a common knowledge representation. In this paper, we present the concept of an information collage (IC) -- a web browser extension combining manual spatial organization of gathered information fragments and automatic text analysis for interactive content exploration and expressive visual summaries. We used IC for case studies with knowledge workers from different domains and longer-term field studies over a period of one month. We identified three different ways how users collect and structure information and provide design recommendations how to support these observed usage strategies. 
\end{Abstract}

\section{Introduction}

Many knowledge workers need to acquire and make sense of a lot of information from various sources to guide their creative processes. 
Examples are journalists who need to research background information around which they construct their newspaper story. 
Similarly, consultants need to understand their costumers' problems, and extract information from existing recommendations and guidelines to come up with specific design guidelines for their costumers. 

\emph{Personal information management} (PIM) supports users with the acquisition, the collection, and retrieval of information \cite{jones_personal_2007}, but PIM tools usually do not provide rich visual interfaces for making sense of the collected information. On the other hand, visual analytics systems use computational analysis and interactive visual exploration techniques to support users analyzing large document collections (e.g., \cite{wong_-spire_2004,endert_semantic_2012, fried_maps_2014, kim_topiclens:_2017, herr_hierarchy-based_2017}) but they usually provide only very limited possibilities to manually structure the contained information. Knowledge externalization strategies, such as concept maps or mind maps, are powerful approaches to offload cognitive effort \cite{scaife_external_1996} and supporting cognition by visually perceiving the externalized information \cite{zhang_nature_1997,kirsh_computational_1995}. However, knowledge externalization tools are usually decoupled from collected information. 

The information collage (IC) bridges information acquisition, collection, analysis, and knowledge externalization. This environment allows users to collect and structure information of different types and granularities in a ``collage''. The concept of IC is inspired by fine arts, where a collage is defined as \emph{``a unified composition from abstracted fragments of those objects depicted''} \cite{pipes_foundations_2003}. Analogously, the goal of an information collage is to depict the user's knowledge by fragments of information from various sources, combined with own thoughts and interpretations. The collage is enriched by computational analysis of collected content, which is used to support semantic zooming in large information collections and interactive exploration of the content. Using IC, we performed case studies with four knowledge workers with different backgrounds and received feedback and activity logs from eight longer-term users. 

In summary, our contributions are two-fold: 
\begin{enumerate}
    \item The design of a scalable information collage with computational analysis support, implemented as extension to a web browser. 
    \item The results of case studies and longer-term field studies revealing different strategies how users collect and structure information in such a collage interface. 
\end{enumerate}

\section{Related Work}
\label{sec:relatedWork}

After analyzing knowledge- and information-intensive work, Pirolli and Card \cite{pirolli_sensemaking_2005} grouped the work flow 
of intelligence analysts into two sets of processes: \emph{information foraging} and \emph{sensemaking}. 
Information foraging describes the information seeking, assessment, and handling processes \cite{pirolli_information_1995}. 
Many of these processes are nowadays conducted using online resources, such as web sites or personal data stored in the cloud. 
To cope with the massive amount of online information, users adopted different strategies to investigate multiple online information sources, such as opening information sources in separate browser windows or tabs \cite{aula_information_2005} or copying URLs into dedicated documents \cite{jones_keeping_2001}. 
The usage of bookmarks, however, has been found to be surprisingly low \cite{obendorf_web_2007}; presumably because they are hard to organize \cite{aula_information_2005}. 


Many research prototypes and commercial tools (like Evernote\footnote{https://evernote.com/} or Diigo\footnote{https://www.diigo.com/}) have been developed to support users to extract and structure content from online resources. 
These tools allow users to organize information either into a folder hierarchy or by using tags. Both approaches have their advantages and disadvantages: while folders are somehow more easy to handle, tagging supports more flexible many-to-many mappings \cite{civan_better_2008, bergman_folder_2013}. However, Civan \etal~\cite{civan_better_2008} argue that both approaches are insufficient:
\begin{quoting}
``People think of their information in ways that go well beyond the representational ability of either folders or tags. [...] The ultimate model of information organization may be neither `place this' nor `label this', but instead, `this is how I see things'.''
\end{quoting}

An alternative to strict hierarchical organization or tagging is inspired by organization strategies at physical workspaces \cite{malone_how_1983,kidd_marks_1994}. In freeform spatial interfaces, users can externalize their internal categorizations through unconstrained spatial arrangements. 
In TopicShop \cite{amento_topicshop:_2000}, Session Highlights \cite{aula_information_2005}, or Data Mountain \cite{robertson_data_1998}, for instance, bookmarked web sites are shown as thumbnails that can be freely positioned on a two dimensional plane or tilted surface. 
The Sandbox \cite{wright_sandbox_2006} provides similar capabilities within an information retrieval and analysis environment, but additionally supports sophisticated annotation. 
The ScratchPad \cite{gotz_scratchpad:_2007} is a web browser extension that allows users to organize snapshots of images, text fragments, or complete web pages spatially in an information graph. 
Similarly, in Sensemap \cite{nguyen_sensemap:_2016} users can add web sites from a comprehensive browsing history as nodes to a freely arrangeable knowledge graph. 
IC supports free spatial organization of information fragments. In contrast to these examples, it can scale to a large number of collected information fragments and supports in-depth content exploration by using natural language processing. 

The majority of information knowledge workers are dealing with today is text-based. 
Making sense of text-based information has been explored extensively in the visual analytics community. 
A prominent example is Jigsaw, which visualizes entities extracted from a document corpus in multiple coordinated views \cite{stasko_jigsaw:_2008}. 
Besides standard visualizations, like scatter plots and graphs, users can also freely arrange, group, and link documents in a ``shoebox'' view.  
Corpus visualization techniques, like IN-SPIRE \cite{wong_-spire_2004} or MoCS \cite{fried_maps_2014}, visualize documents mapped onto 2D space based on their word and phrase similarities. 
Since such 2D projections can become very cluttered, corpus visualizations are sometimes constructed hierarchically, where labels are assigned to aggregations of documents depending on the zoom level \cite{andrews2002infosky, paulovich2008hipp, endert2013typograph, herr_hierarchy-based_2017}. 
Others use a magic lens metaphor to explore more detailed topic clusters in a large document collection \cite{heimerl2016docucompass, kim_topiclens:_2017}. 
IC also uses representative keywords to create expressive summaries, but clusters are derived from the user's spatial organization. In addition, users can further manually structure the collage using notes, highlights, or explicit groupings. 


Corpus visualization techniques, as described above, compute the layout of documents automatically based on content similarity. However, this may not correspond to the user's mental model of what content belongs together. On the \emph{Analyst's Workspace} \cite{andrews_analysts_2012}, users can therefore manually organize text documents on a large display. 
Entities are automatically extracted and visual links reveal shared entities across the text documents. 
Others automatically compute a layout for documents, but allow users to adapt the underlying similarity model \cite{endert_semantics_2012, brown2012dis} or steer the topic modeling technique \cite{choo2013utopian} to interactively change the spatial layout. 
Our IC prototype provides free spatial arrangement of a large number of information fragments captured online. Using IC, we could study two aspects that differ from the examples above: First, the information was captured by the users themselves and is therefore not entirely unknown as compared to a task, where users have to study the content of an entirely unknown corpus. Second, due to the longer-term usage, we could also qualitatively explore the effects of spatial organization for retrieval of already visited information. 

Manual spatial organization is not only useful during information foraging and sensemaking, but also for generating new ideas. 
Post-its are therefore often used to support brainstorming sessions.
This physical metaphor has been adopted by idea generation tools like NoteCards \cite{halasz_reflections_2001} or collaborative brainstorming environments for interactive tabletops \cite{hilliges_designing_2007}. 
Spatial hypertext systems, like AquaNet \cite{marshall_aquanet:_1991} or VIKI \cite{marshall_viki:_1994} support knowledge structuring tasks in the browser, where objects or nodes hold content, such as text or a URL, which can be connected by links, and the system automatically interprets composite structures. 
More classic knowledge externalization tools are mind maps or concept maps \cite{davies_concept_2010, meisterlabs_mindmeister:_2016,christian_foltin_main_2016}. Hierarchical, zoomable mind maps have been shown to improve the user experience \cite{haller_imapping:_2010}.
IC supports similarly flexible knowledge structuring, but enriches this information collage with interactive visual content exploration.

\section{Information Collage}
\label{sec:IC}

IC provides a unified environment, where collected information of various formats and arbitrary online sources can be flexibly structured. IC provides mechanisms to extract three types of information fragments: text passages of varying length (Figure \ref{fig:snippets} top), entire online documents, or images. Captured information can be enriched by personal notes. 
To avoid information fragmentation, IC is designed as persistent information environment, where users can collect a variety of information that is not limited to a single topic, project, or task. IC can make use of a large screen space to facilitate information structuring through spatial organization. The structuring of information mimics physical information collections, while IC adds computational power to overcome some of their shortcomings.

\begin{figure}[h]
 \centering
   \includegraphics[width=3.1in]{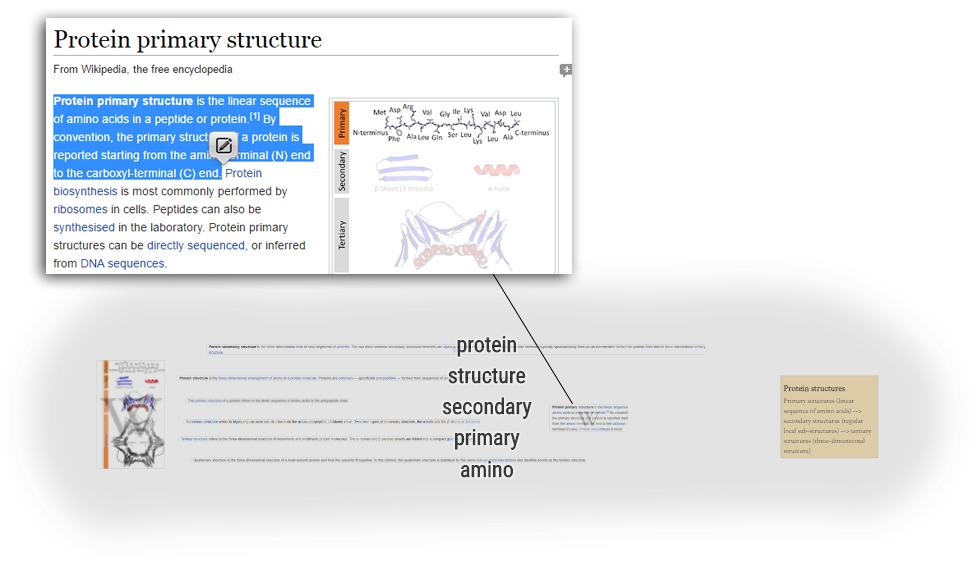}
  \caption{Users capture text or images directly from the web sites (top) and arrange these information fragments spatially in the information collage (bottom). Collected information can be enriched with ``digital post-its'' (right). }
  \label{fig:snippets}
\end{figure}

\subsection{Spatial Organization}

The collage interface is displayed in a separate browser tab. 
Here, all gathered information fragments are provided in an inbox list, chronologically sorted by capturing sequence. Information fragments are represented by a screenshot of the captured item (Figure \ref{fig:snippets} bottom), or the visible portion of a whole document. From the inbox list, the users can drag the information fragments into the collage area and arrange them spatially.

The collage area itself is a zoomable space providing the user opportunity to apply informal spatial structuring of collected information, like using labeled paper cards. According to Gestalt theory, spatial proximity indicates grouping of items. Manual spatial arrangement of collected elements requires reflexion and therefore improves understanding \cite{kandogan_how_2011} and memorability of the location for more efficient retrieval \cite{robertson_data_1998}. In addition, users can add ``digital post-its'' to link the extracted information with their own thoughts or interpretations (Figure \ref{fig:snippets} right). Information fragments and notes can also be grouped into colored and labeled containers to create visually distinct sub-collections. To support fluid switching between the user's information collection and the original information, every fragment is linked to its source. By selecting the fragment, users have the option to return to the origin document, and the exact source location within this document, respectively. 

\subsection{Semantic Zooming}

\begin{figure*}[tb]
   \includegraphics[width=\textwidth]{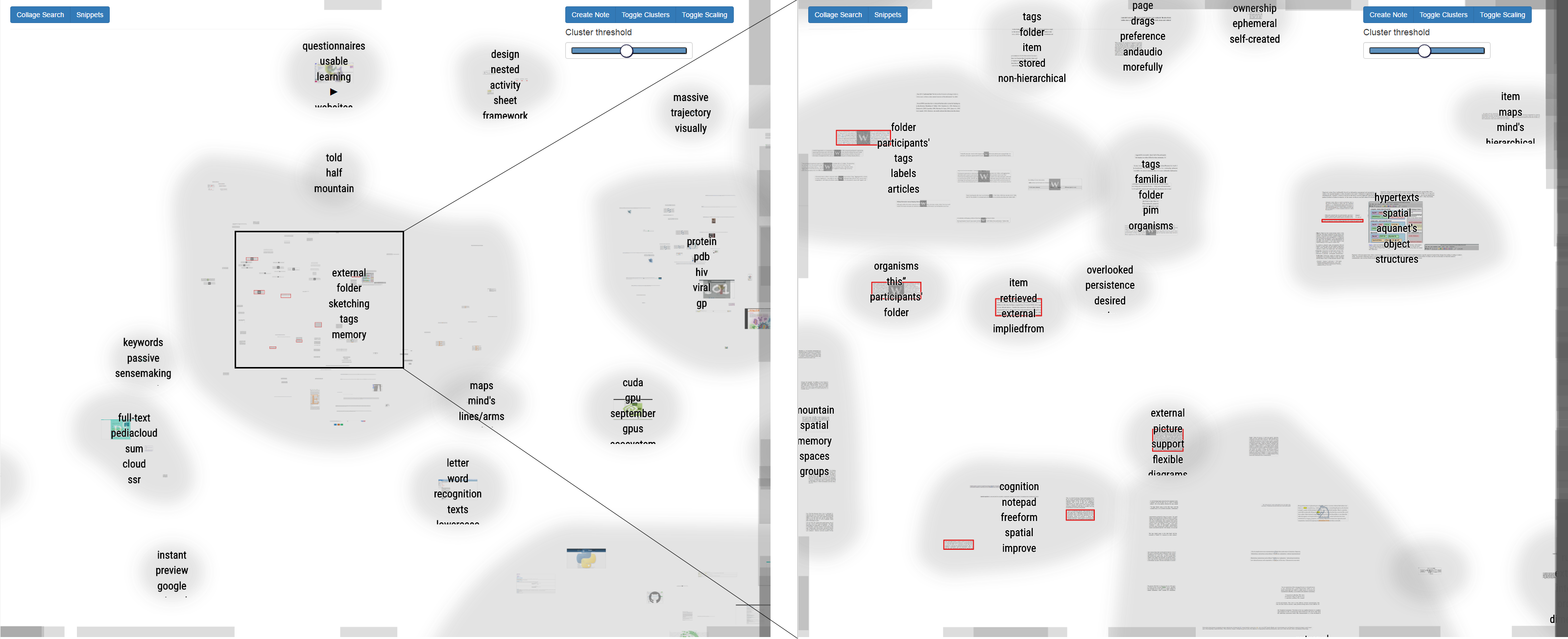}
  \caption{Zooming in a large collage containing citations from different research fields. The aggregated cluster on the left is split up into several sub-clusters. Semi-transparent ``citylights'' at the borders indicate clusters outside the viewport. }
  \label{fig:zoom}
\end{figure*}

A single persistent information collection can quickly reach a size that goes beyond the physical counterpart of a table with clusters and piles of physical paper. When zooming out to get an overview of the collected information, the gathered information soon becomes unreadable. As a consequence, structuring and relocating information becomes a challenge. 

Aggregating and visually abstracting the content when zooming out is therefore a common strategy in zoomable interfaces \cite{bederson_pad++:_1994} and visualizations \cite{elmqvist_hierarchical_2010}, also denoted as \emph{semantic zooming} \cite{perlin_pad:_1993}. 
As the user's structure of the information space is expressed spatially in IC, the system clusters information fragments based on their spatial proximity using a density-based clustering. Fragments being closer to each other than an adjustable distance threshold are assigned to the same cluster. This distance threshold is defined in screen space.
This means, when zooming out, the information density in the clusters increases. This also means that clusters always occupy approximately the same amount of screen space.  
This way, users can structure collage elements into topics and sub-topics by proximity, which are gradually revealed when zooming in (Figure \ref{fig:zoom}). 

To expressively summarize the content, we analyze the information fragments' text content and the users' annotations. For each cluster, we super-impose the five most relevant keywords, which fade out when the user zooms in close enough for the content to be readable. 
Fragments themselves cross-fade to the web page's favicon when zooming out and the text is no longer readable. Clusters outside the visible viewport are indicated by \emph{citylight} off-screen visualizations \cite{zellweger_city_2003}.  

In the zoomable collage interface, users can freely zoom and pan to gradually explore their information collection. On the overview level, the entire collage shows only a few main clusters, separating, for instance, a work-related information collection from private information, or information gathered for different projects. When zooming in, clusters split into finer-grained topics, until a cluster represents only a single information fragment, or overlapping fragments. 

\subsection{Interactive Exploration}
\label{sec:interactiveExploration}

While users can arrange information based on their own mental model, the information fragments may also contain relations that are not immediately evident or relevant to the user. In an intelligence analysis task, Goyal \etal~\cite{goyal_effects_2013} showed that users were more successful extracting a crime plot out of a corpus when provided with a visualization of shared terms between documents. 
We therefore use the fragments' text content to compute the similarity of all collage items to a selected item based on their shared terms. A selected item can be a single information fragment or note, a cluster, or an information fragment located in the inbox list. 
Showing the similarity between an information fragment and the inbox and the collage fragments also serves as visual guidance to place new information close to similar information in the collage. 

The clusters' similarities to the selected item are visualized by opacity: the darker a cluster, the more similar to the selected item, as shown in Figure \ref{fig:similarity}. This encoding is not very accurate, yet it helps the user to quickly identify which collage fragments share text content with the current selection. In addition, we show up to two shared keywords to not only visualize \emph{that} there is a similarity, but also \emph{how} the items are connected. To reveal the role of the keywords in the respective items, \emph{keyword-in-context} (KWIC) boxes show information fragments' text containing the keyword when hovering it. 

\begin{figure}[tb]
 \centering
   \includegraphics[width=0.7\textwidth]{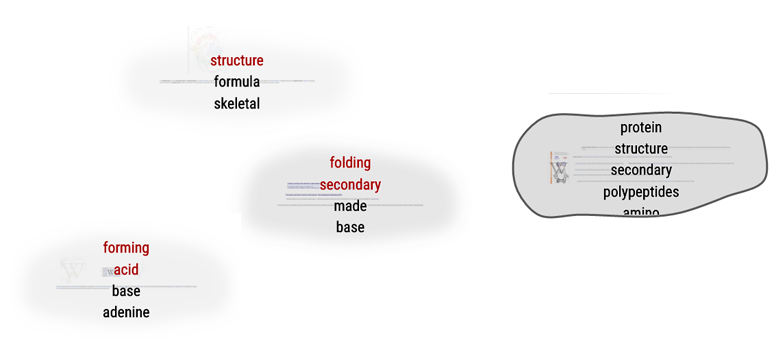}
  \caption{Selecting an aggregated cluster of information fragments (right) reveals similarities to other clusters: red terms are shared with the selected cluster, and overlay opacity visualizes text-based similarity. }
  \label{fig:similarity}
\end{figure}

\subsection{Search for More Information}

Search for additional information can happen at any time during a knowledge-intensive task. In addition, as users consume more information, they gradually rephrase their query terms \cite{bates_design_1989}. 
We support this process by providing a context menu option to search for information related to a selected cluster on the web. The query is constructed by the five visible cluster keywords and is sent to the Google search engine in a new browser tab.

\subsection{Implementation}

IC is implemented as extension to the currently most popular web browser Google Chrome \cite{wikipedia_usage_2017} in JavaScript.
It utilizes the D3 library \cite{bostock_d3:_2011} for the SVG-based collage visualization, the Annotator.js library\footnote{http://annotatorjs.org/} for highlighting and annotating web site passages, and PDF.js\footnote{https://mozilla.github.io/pdf.js/} for transforming PDF documents into HTML, to allow information capturing from online PDF resources. The data layer of our application is backed by IndexedDB, a client-side storage for quick access to large data collections.


Information fragments are spatially clustered using the jDBSCAN\footnote{https://github.com/upphiminn/jDBSCAN} library based on the two closest points between pairs of fragments. The distance threshold for the clustering is zoom-dependent. The graphical representation of the computed clusters is created by computing the concave hulls for the set of rectangles in each cluster and drawing a catmullRom spline around these points. We create the concave hull with the hull.js\footnote{https://github.com/hull/hull-js} algorithm. 

When the user creates a new text fragment or captures an entire web page or converted PDF, the contained text is extracted and the following text processing steps are performed to compute weighted keywords for each fragment: 1) stop word removal, 2) stemming of terms using the jsSnowball\footnote{https://github.com/fortnightlabs/snowball-js} library, 3) computation of term frequencies (\emph{tf}), 4) update of the global inverse document frequencies (\emph{idf}) of these terms (where ``document'' corresponds to information fragment here), and 5) (re)computation of \emph{tf*idf} \cite{manning_introduction_2008} weights for all terms in all fragments. Term frequencies are also computed for all current clusters by summing up the containing fragments' tf-values. To identify expressive cluster terms, tf*idf weights for all terms in a cluster are computed (where ``document'' corresponds to cluster here). 
The five terms with the highest tf*idf weights are used to label the clusters. 
For interactive exploration of fragment similarities, we use the cosine similarity measure \cite{manning_introduction_2008} to quantify the similarity between a selected information fragment, note, or cluster to all other clusters in the collage. 

\section{Evaluation}

Evaluating users' knowledge work flows is, in general, challenging. Tasks are often performed over a long period of time and can be frequently interrupted \cite{czerwinski_diary_2004} and resumed at a later point. Also, knowledge work can have very diverse goals, yet similar work flows or -- conversely -- very distinct work flows to reach similar goals. Most importantly, the usage of IC has to be analyzed in the context of the user's normal work environment to be able to characterize the interplay between online information, search engines, and other applications in the user's work flow.  
A classic short-term evaluation in a laboratory setting is therefore insufficient to provide detailed insights how people collect and structure information in a collage interface. 

We therefore chose to conduct case studies with knowledge workers from different domains to be able to observe how IC could fit into their knowledge-intensive work flow. To be able to observe revisitation and spatial organization of collected personal information over a longer period of time, we asked 18 computer scientists to install and use IC over a period of one month. 
Our evaluation was split into three blocks of studies using different versions of the IC prototype. Figure \ref{fig:method} shows the temporal sequence of studies and the IC prototype versions used for the respective studies. 

 \begin{figure}[h]
 \centering
  \includegraphics[width=0.8\textwidth]{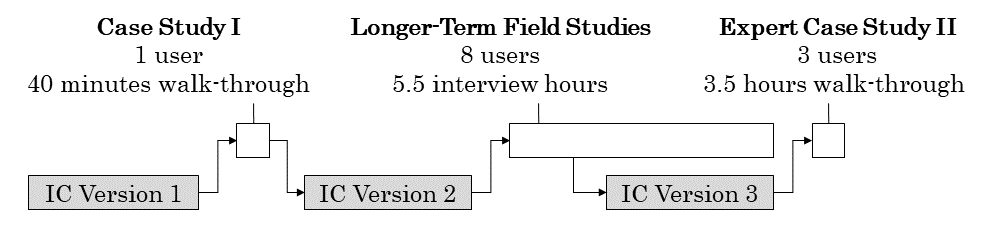}
  \caption{Temporal sequence of our studies and IC prototype versions. }
  \label{fig:method}
\end{figure}

\subsection{Information Collage Versions}

As we iteratively incorporated feedback from the users into the IC prototype, the studies were conducted with different prototype versions: Version 1 supported spatial organization of information fragments, but did not feature semantic zooming. Interactive exploration was supported through term boxes attached to selected information fragments, and visual links were used to show the most similar fragments to a selected fragment (Figure \ref{fig:IC_V1}). In addition, this version featured an integrated web search engine based on Google search that colored search hits based on their similarity to the collage content, and also provided term boxes of keywords shared with collage content, as well as the most relevant unique keywords. 

 \begin{figure*}[htb]
 \centering
    \subfigure[Version 1]{
        \includegraphics[width=0.3\textwidth]{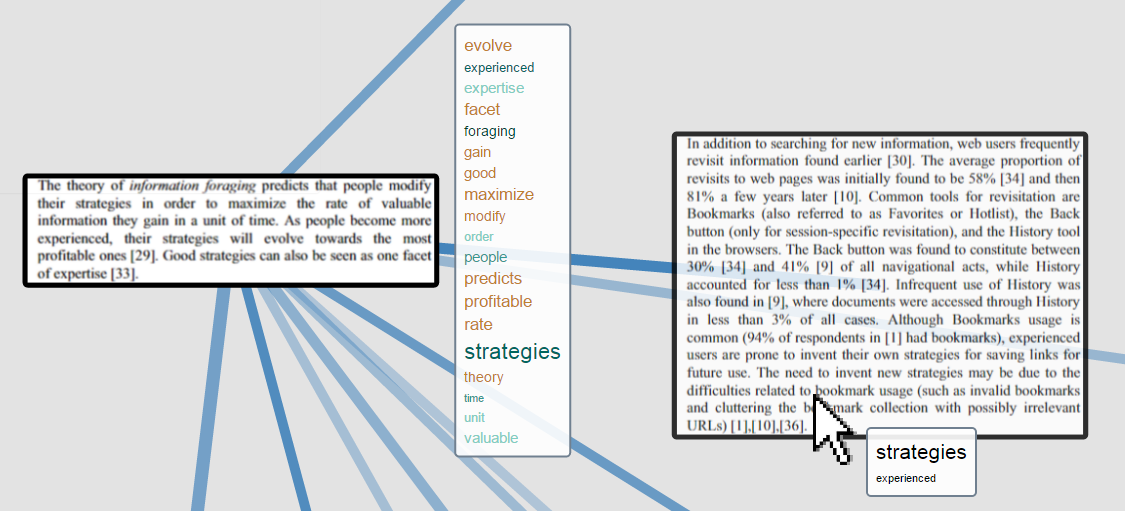}
        \label{fig:IC_V1}
    }
    \subfigure[Version 2]{
        \includegraphics[width=0.3\textwidth]{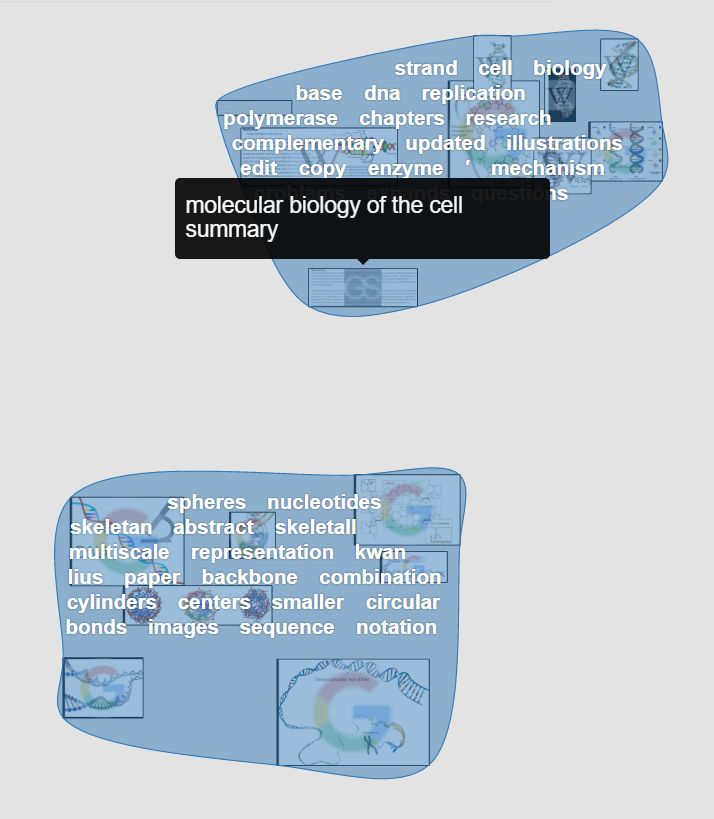}
        \label{fig:IC_V2}
    }
    \subfigure[Version 3]{
        \includegraphics[width=0.3\textwidth]{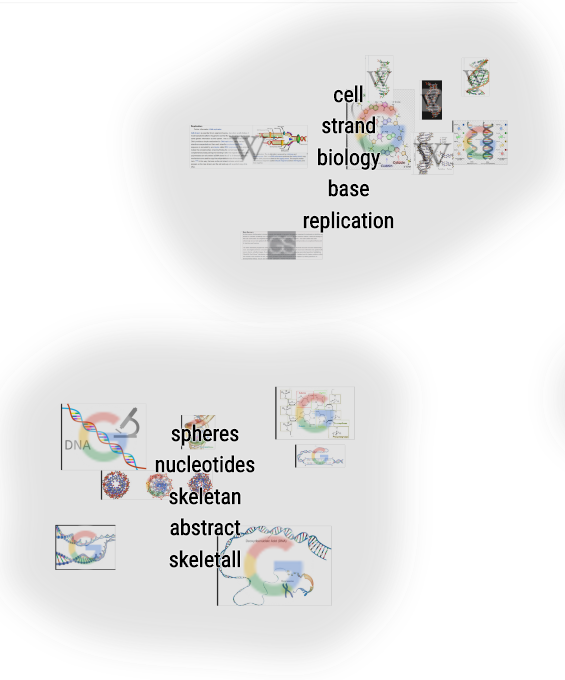}
        \label{fig:IC_V3}
    }
    \caption{The three versions of IC: the first version used visual links to indicate similarities and showed a list of shared (blue) and unique (brown) keywords (a), the second version introduced clustering instead of visual links (b), and the final version (showing the same collage as in (b)) used a different visual design (c). }
    \label{fig:IC_versions}
\end{figure*}

Version 2 was created after the first expert case study with a journalist. We eliminated the integrated web search engine, because the user preferred the look-and-feel of her familiar search interface, and also missed crucial search features, like image or news search. Visual links and term boxes were eliminated because of visual clutter, even in sparsely populated collages, and substituted by the semantic zooming and cluster similarity exploration techniques, as described in Section \ref{sec:interactiveExploration}. Compared to the final Version 3 (Figure \ref{fig:IC_V3}), Version 2 showed more cluster keywords and had a different graphical design (Figure \ref{fig:IC_V2}). 
We also fixed smaller bugs and usability problems in Version 3 that were reported during the longer-term case studies. 

\subsection{Case Study Design}

For the case studies, we asked four knowledge workers (two females) from different domains to walk us through a typical knowledge-intensive task with the help of the IC prototype. We were interested in their elicited use cases, as well how well the information collage concept fits into their work flow. Users were asked to select a representative use case, perform the task with the support of IC, and to give feedback. 

The case studies were performed on one of the researcher's PC. We performed case study I with a journalist using IC version 1. Case study II was conducted with a content-experience designer, an experience strategist, and a video producer using IC version 3. All walk-throughs were audio-recorded and transcribed, and all resulting collages were stored. 

\subsection{Longer-Term Field Study Design}

For the longer-term field studies, we asked 18 volunteers with a computer science background (seven undergraduate students, five graduate students, four research scientists, and two software engineers; two females and 16 males) to install IC on their PCs. They were asked to use IC whenever they thought they could benefit from it within a period of one month. Again, we were interested in their use cases, but also in how frequently they would access the collage and revisit their stored information sources, as well as how they would benefit from IC's capabilities to navigate and explore large information collections. We introduced users to the capabilities of IC and asked them to perform typical information- and knowledge-intensive and / or creative tasks using IC. However, we did not prescribe any usage scenarios. 

We logged all IC activities over a period of approximately one month. In addition, we also obtained qualitative feedback through interviews one to two weeks before finishing the study. From the initial 18 users, we received feedback from 14 users. Three of these users did not use IC within the study period, and three users just \emph{``played around with it''}, but did not perform an actual task. We therefore only analyzed the material of those eight participants who performed work-related tasks with IC. Since activity logs of the longer-term case studies were anonymous, we identify users here by the first two letters of the automatically generated ID of the IC Chrome browser extension.

\subsection{Results}

Grounded in the abstracted work flow depicted in Pirolli and Card's sensemaking loop \cite{pirolli_sensemaking_2005}, we analyzed the activity logs, the interview transcripts, the resulting information collages, and the field notes of the expert case studies to identify  the users' strategies how to integrate IC into their knowledge-intensive workflow. Table \ref{tab:cases} lists all users with their reported use cases.

\begin{table}[h]
\begin{center}
\scalebox{0.9}{
\begin{tabular}{| l | l |}
 \hline
 \textbf{user} & \textbf{use case}  \\ \hline
 journalist (J)						& article background 			 \\
  \hline			
  experience strategist (E) 			& content management tools 		 \\
  
  content-experience designer (C)		& design guidelines				\\
  video producer (V)					& purchase decision and movie script 			 \\  \hline
                                   
  KO 								& related work structure 			 \\
  LP 								& ``internal brainstorming''			 \\
  EM								& state-of-the-art report  \\
  PG 								& literature research			 \\
  EG 								& literature research			 \\
  DA 								& literature research			 \\
  CE 								& bookmarking			 \\
  LN 								& bookmarking			 \\
  \hline  
  
\end{tabular}
}
\end{center}
\caption {List of users of case studies I and II (upper rows) and the anonymized users of the longer-term field studies (lower rows) and their use case(s). }
\label{tab:cases}
\end{table}

\subsubsection{Observed Work Flows}
\label{sec:workflows}

All four professionals in the case studies started with an online \emph{search} and opened multiple browser tabs with potentially interesting documents. They then started to skim or \emph{read} the information in multiple closely related documents and to selectively \emph{extract} information or \emph{store} the entire document. After having spent some time on browsing web sites on the same topic, they switched to the collage. There, they first coarsely positioned the information fragments before searching for more information in their preferred search engine. Usually, information fragments from one such ``swipe'' were initially just placed closely together. This was repeated two or three times, before they switched to the collage to \emph{structure} their information. 

While the general embedding of IC into the information foraging process, as described above, was very similar among all four professionals, the further processing of the information within IC differed between the users. To differentiate and characterize user strategies, we present the analysis of the activity logs of a larger group of users in the longer-term field studies in Section \ref{sec:usage}. 

\subsubsection{Analysis of Collages}

We categorized all information fragments in the resulting collages into pure image fragments, information fragments containing only a few words to a few lines of text, complex snippets containing either very long text fragments or combining multiple content types, like tables, images, and text, whole documents, and user-created notes. 
As shown in Figure \ref{fig:snippetTypes}, most users preferred a single fragment type, where short texts were most common, followed by whole documents. User LP was the only user relying mainly on images. In the interview, he explained that he was researching different ways how to visually depict DNA, which explains why most material he collected were images. The number of created notes was generally very low. One notable exception is user KO, who applied very extensive manual structure onto his collage for the related work research of his thesis. 

 \begin{figure}[h]
 \centering
   \includegraphics[width=0.6\textwidth]{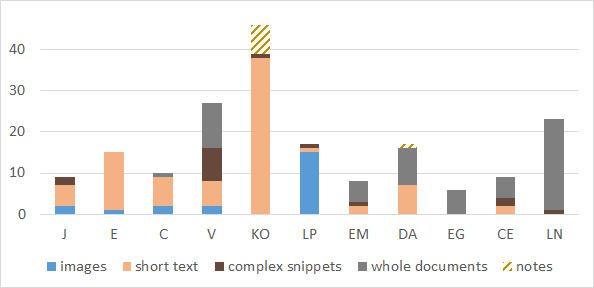}
  \caption{Number of information fragments in the final collage for each user, grouped by fragment type (user PG did not share his collage). }
  \label{fig:snippetTypes}
\end{figure}

Of the eleven analyzed collages, seven were focused on a single topic, such as related work for a current research project (see Table \ref{tab:cases}). In one of the case studies, the video producer used a single collage to analyze video hardware to make an informed purchase decision, as well as to research background information for a science movie script. Two users of the longer-term field study [CE, LN] collected information about multiple topics, including both, work-related and private information (see, for instance, Figure \ref{fig:bookmarks}). 

\subsubsection{Analysis of Activity Logs}
\label{sec:usage}

For the users in the longer-term field studies, we also analyzed how many information fragments they created in total (including those that were later deleted), how often they visited the collage during the study period, and how often they revisited a source document through a collage fragment. Figure \ref{fig:fieldLogs} shows these logs for the eight users. 

 \begin{figure*}[htb]
 \centering
    \subfigure[]{
        \includegraphics[width=0.3\textwidth]{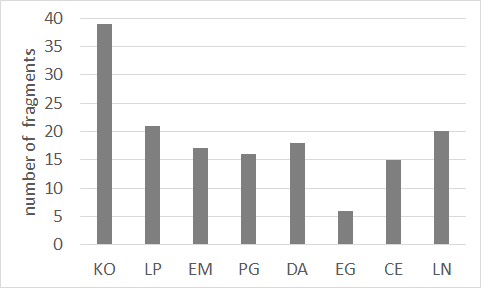}
        \label{fig:numSnippets}
    }
    \subfigure[]{
        \includegraphics[width=0.3\textwidth]{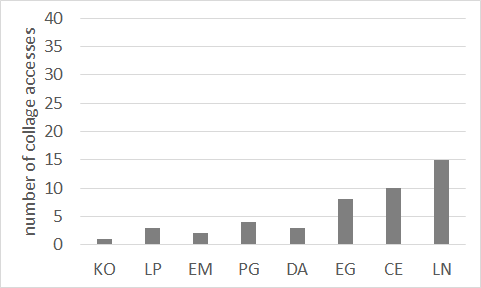}
        \label{fig:numAccess}
    }
    \subfigure[]{
        \includegraphics[width=0.3\textwidth]{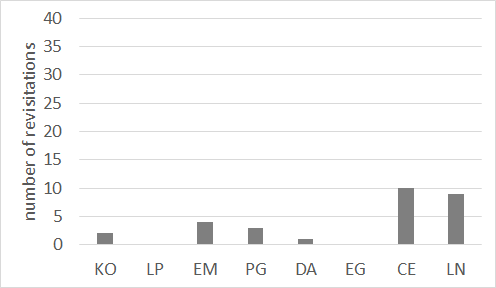}
        \label{fig:numRevisitations}
    }
    \caption{Field logs of the longer-term case studies: (a) number of information fragments created, (b) number of collage accesses, and (c) number of source revisitations. }\label{fig:fieldLogs}
\end{figure*}

We standardized the three-dimensional attribute space depicted in Figure \ref{fig:fieldLogs} by centering the eight samples and by scaling the axes to unit variance. We then clustered the samples in this standardized three-dimensional attribute space using density-based spatial clustering with a maximum distance of 2.0. This way, we obtained the following three clusters: [KO], [LP, EM, PG, DA, EG], and [CE, LN]. We qualitative analyzed the activity logs, the resulting collages, as well as the interviews to characterize these activity-based clusters and derived the following three usage strategies, described in more detail below: 

\begin{itemize}
    \item information and knowledge map (IKM), 
    \item shoebox (SB), and
    \item personal information collection (PIC). 
\end{itemize}

To create an \emph{information and knowledge map}, one user [KO] carefully selected information on a single topic and applied detailed manual structure, such as outlines, headlines, and notes (see Figure \ref{fig:paperStructure}). His collage contained a high number of mainly text-based fragments, but after finishing his knowledge map, he only returned to the collage a single time within the logged time period. He also revisited source information only very rarely. 

 \begin{figure}[h]
 \centering
   \includegraphics[width=0.7\textwidth]{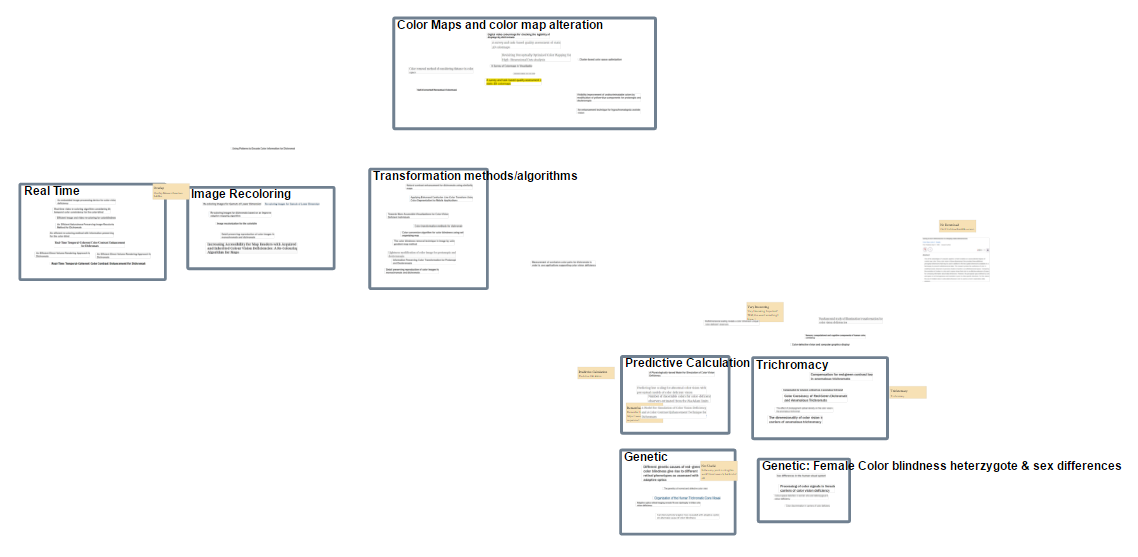}
  \caption{Example for an information and knowledge map: structure of a related work chapter with titles of papers grouped into labeled sections (user KO). }
  \label{fig:paperStructure}
\end{figure}

 Most users [LP, EM, PG, DA, EG] exhibited a strategy that can most closely be described by Pirolli and Card's notion of a \emph{shoebox} \cite{pirolli_sensemaking_2005}: instead of opening all web documents that could be interesting for the present task in new tabs, they stored entire web documents or long text passages from web sites, images, PDFs, or e-mails to the collage for later reading (see Figure \ref{fig:literature}). Inside the collage, they only coarsely grouped the information fragments. These users mostly reported that the fully manual spatial organization was considered as unnecessary overhead for them. Some users rather requested automatic organization of the collected documents based on the content. Furthermore, they expected to be able to read and filter the collected information inside the collage after having stored a (temporarily) sufficient amount of source material. 
 
 \begin{figure}[h]
 \centering
   \includegraphics[width=0.7\textwidth]{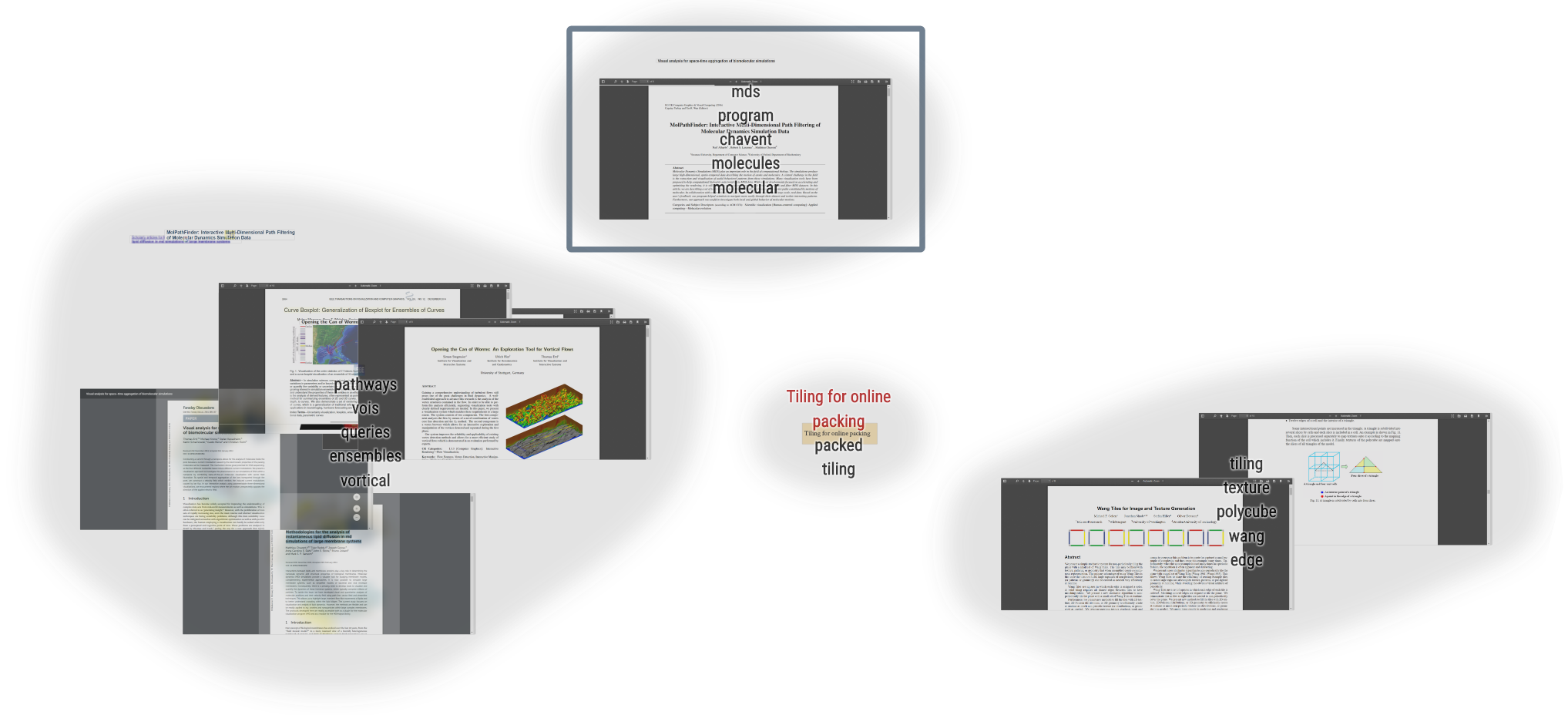}
  \caption{Example for a shoebox: PDF snapshots of papers on a particular topic (user DA).  }
  \label{fig:literature}
\end{figure}

Two users [CE, LN] started to use IC as a \emph{personal information collection (PIC)}, where carefully selected references on various topics with varying degrees of relatedness were roughly positioned in the collage for later retrieval. Their collages contained entire web documents, but also some longer text passages were stored. In contrast to the shoebox strategy, their collages covered multiple topics with varying levels of similarity between the information fragments. Spatial proximity was used to express topic associations on multiple levels. User LN described his organization strategy as follows (\cf, Figure \ref{fig:bookmarks}):
\begin{quoting}
``Bitcoin-related information belongs together. Programming documentation is close to the Bitcoin-things, but not too close. My free notes are very far away.''
\end{quoting}
Spatial organization thereby primarily served as memory aid to be able to quickly find a specific piece of information again. Compared to the other users, PIC-users had a relatively high number of collage accesses and a high number of source revisitations. 

\begin{figure}[h]
 \centering
   \includegraphics[width=0.6\textwidth]{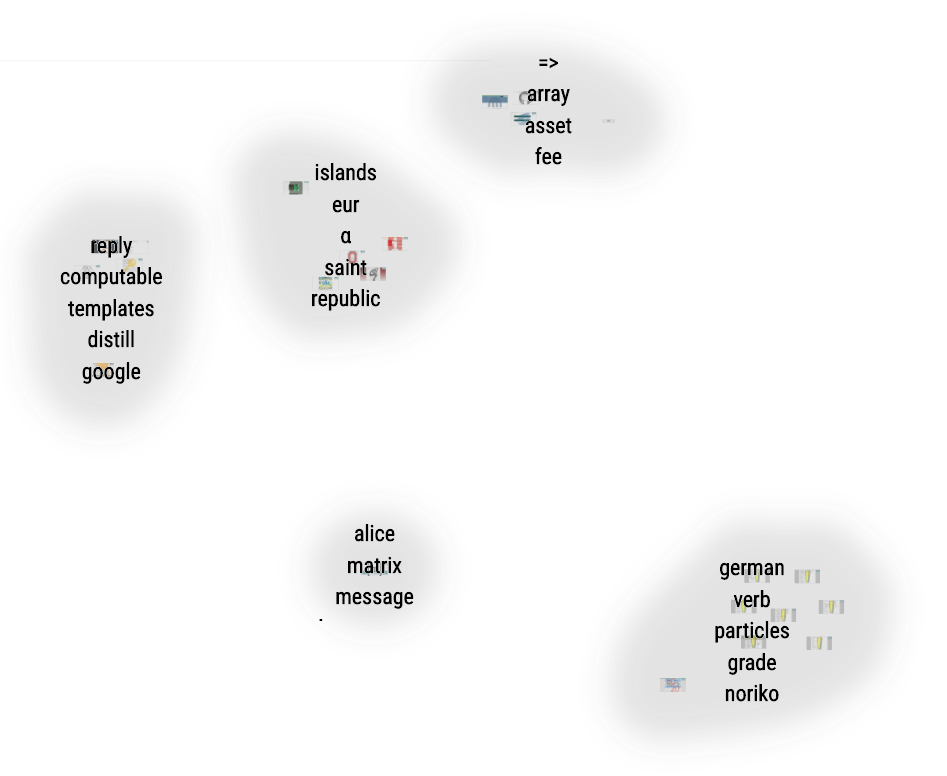}
  \caption{Example for a personal information collection: bookmark piles for different work-related topics (top left) and a private (bottom right) topic (user LN). }
  \label{fig:bookmarks}
\end{figure}

The usage patterns qualitatively observed during the case studies mostly fit to the SB-strategy. The only exception was the experience strategist, who created a carefully organized information and knowledge map of content management tools. We could not observe that case study users started to create a personal information collection in the short study duration.

\subsection{Strengths and Limitations of the IC Concept}

Below, we discuss the major design aspects of IC with respect to our observations and the gathered user feedback. 

\subsubsection{Spatial Organization and Information Structuring}

The most apparent feature of the information collage for the users was the ability to freely organize collected information fragments in 2D space. Even the knowledge workers from outside the computer science domain immediately understood this interaction concept and could easily apply it to organize collected information. What differed between the observed usage strategies was what the users aimed to express using the spatial organization: While the IKM-user utilized the space to structure collected information fragments into meaningful categories, PIC-users made use of spatial memory to easily find stored references again. 
SB-users employed the information collage for potentially relevant information sources without intending to spend effort to organize it. Having to place the captured material manually was considered too tedious for many users. It can also be expected that, after storing a whole document without prior inspection, they were not confident enough about the content to be able to properly categorize the document. 

Apart from spatial organization, IC provides some other methods how to structure and enhance the collected information, such as by explicit grouping, highlighting important information fragments, and adding notes. Participants made little use of these features. Exceptions are IKM-users, who expressed the wish to have even more control over the visual appearance of information. They requested additional features like manual resizing of collage elements and color-coding of individual information fragments and clusters. 
Compared to a classic mind map, these users appreciated that the spatial organization in the collage is \emph{``free and unconstrained''}, and that they have the option to easily go back to their original information sources, which makes the \emph{``research path extremely easy to retrace.''} 

\subsubsection{Semantic Zooming}

The ability to zoom the collage was, in general, appreciated by users. The keyword overlays, however, intending to summarize the underlying information, lacked the expressiveness to be truly useful for the participants. While the keywords helped to \emph{``somehow know what [the cluster] is about''} (content-experience designer), clusters often contained irrelevant or incorrectly extracted keywords so that users were not able to reliably infer the content underneath. One user stated that \emph{``it does not find the relevant terms but only frequent ones''}  -- an observation shared with Endert \etal~\cite{endert_semantics_2012} for intelligence analysis documents. Users therefore sometimes preferred to group and label their clusters manually. 
Figure \ref{fig:keywords} shows how the experience strategist labeled groups of information fragments with names of content-management tools, and which keywords were extracted by the system for the same fragments. The experience strategist considered these keywords as \emph{``useless''}. 

\begin{figure}[h]
 \centering
   \includegraphics[width=3.1in]{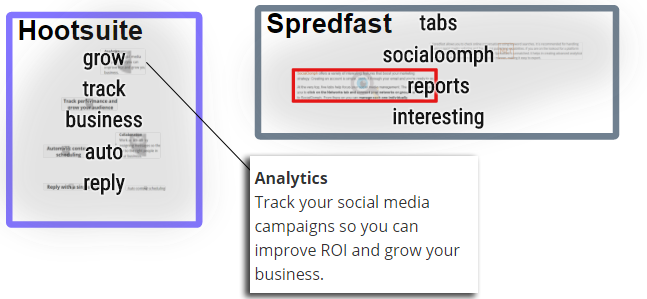}
  \caption{User-generated container labels (names of content-management tools) compared to automatically extracted cluster keywords (experience strategist). The inset shows an enlarged information fragment. }
  \label{fig:keywords}
\end{figure}

IKM- and SB-users particularly disliked the merging of clusters and their keywords, respectively, when zooming out. User EG, for instance, explained: 
\begin{quoting}
``What confuses me is this big groups. For me, that if I collect something together, has a meaning. When I zoom out, [the system] does it automatically.''
\end{quoting}
While IKM-users often switched to manual labeling instead (see Figure \ref{fig:paperStructure} and Figure \ref{fig:keywords}), manual labeling or tagging was considered too tedious when quickly storing information to the collage by SB-users.  

Despite having installed the information collage for a period of one month, many collages were fairly small in the end. Since keyword aggregation for semantic zooming was designed to aid navigation in \emph{large} information collections, the usefulness of this concept may have been underrated by the users. 
Also, similarity exploration was not deliberately used nor commented by the users. However, we cannot conclude here that it was not useful. Rather, the lack of keyword expressiveness may have limited the usefulness of this approach. 



\subsection{Study Limitations}

The IC browser extension was a research prototype and therefore lacked the perfection in terms of usability and performance that users expect from a tool they use on a daily basis. For instance, some users disabled the Chrome extension after a while, because selecting a text on a website would trigger the capturing interface (see Figure \ref{fig:snippets}), sometimes even with slight latency. Also, background parsing of web sites for image capturing sometimes affected the performance of the browser. 
We cannot rule out the possibility that these technical limitations, as well as the lack of familiarity with IC, had an influence on the users' work flows, in particular the usage frequency and the resulting size of the collage. 

\section{Design Recommendations for Information Organization Tools}

Our results show that users apply fundamentally different usage strategies when collecting and organizing their information. Figure \ref{fig:strategies} summarizes these observed strategies. We distinguish between four work steps observed in our case studies (Section \ref{sec:workflows}): \emph{reading} of the content in an online resource, \emph{extracting} relevant information fragments of these online resources or, alternatively, \emph{storing} whole documents, \emph{structuring} these information fragments or documents, \emph{revisiting} the source content associated with the collage content, and \emph{searching} for additional information to extend the collage.

 \begin{figure}[htb]
 \centering
        \includegraphics[width=\textwidth]{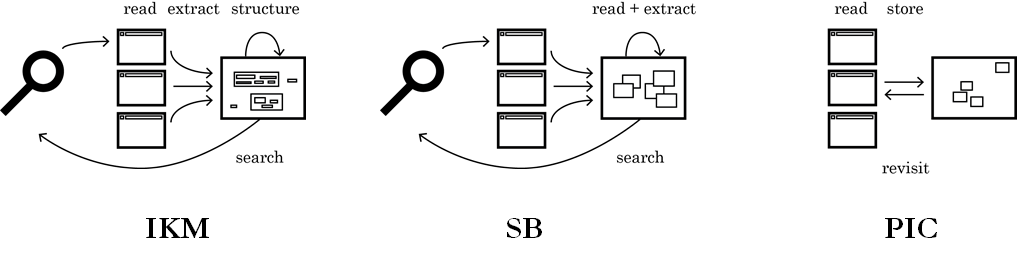}
    \caption{Generalized work flow for the observed usage strategies: extracting small information fragments on one topic and structuring them in the collage (IKM), storing entire online documents and complex snippets on one topic in a shoebox (SB), and selectively storing entire online documents, roughly arranging them topic-wise, and revisting the sources later (PIC). }\label{fig:strategies}
\end{figure}

Figure \ref{fig:strategies} illustrates that the work flows for the observed usage strategies are fundamentally different from each other. It therefore seems unlikely that a single tool can cover all observed usage strategies. Based on our observations and user feedback, we provide design recommendations for information organization tools focusing on either of the three usage scenarios. 

\subsection{Information and Knowledge Maps}

When creating information and knowledge maps from carefully selected information fragments extracted from web pages, users expect to combine these information fragments with their own mental model. They therefore need to express their mental models through a variety of visual channels, such as spatial proximity, visual boundaries around information fragments, colors of boundaries and fragments, as well as text labels. Environments like the Sandbox \cite{wright_sandbox_2006} or spatial hypertext systems \cite{marshall_aquanet:_1991,marshall_viki:_1994} provide such features. As an extension, automatically extracted keywords and similarities could serve as recommendations for labels or color codes, but IKM-users need full control to correct these suggestions.

As shown in Figure \ref{fig:literature}, an IKM can grow quite large. Semantic zooming therefore seems to be crucial. However, visual summaries based on the content of the information fragments, as provided by IC, is not recommended. The users' self-assigned headers and notes can be expected to describe the knowledge space more expressively (see Figure \ref{fig:keywords}). However, the information fragments were appreciated as links back to the original information source by our users . 

\subsection{Shoebox}

The shoebox-strategy was observed most frequently and therefore seems most promising to reach a lot of users. Using this strategy, users search for information and put every potentially interesting document or other larger chunk of information into the collage for later detailed inspection. Some users expected the tool to automatically arrange the captured information based on the content. As the content was unexplored at the time being pushed into the collage, the problem is similar to classic corpus visualization techniques that show similarity-based projections of an unknown document collection (e.g., \cite{wong_-spire_2004,fried_maps_2014,herr_hierarchy-based_2017}). However, a document corpus usually has structured meta-data like authors or publication years that can be utilized to interactively explore the corpus. Such information is often unavailable or irrelevant when capturing information from the web. On the other hand, there is other valuable meta-information, such as capturing time or provenance information (e.g., common referring pages or similar query terms leading to the captured information) that could be considered when computing the similarity between captured documents. 

Mind that IKM-users rather used multiple browser tabs as shoebox, which is a commonly observed strategy in practise \cite{aula_information_2005, nguyen_sensemap:_2016}. They used IC at a later stage, when the already known and filtered information was categorized. It can be expected that shoebox-users will reach the same state at some point, after they extracted relevant, fine-grained information fragments from the material in their shoebox. In our evaluation, we could not observe such a transition as the current version of IC does not support the extraction of information fragments from collage elements. Our SB-users missed options to gradually explore the contained information, extract related information from multiple fragments, compare them, filter them according to different attributes, and re-arrange them according to these attributes. This is a major challenge, since the contained data is highly unstructured, noisy, potentially incomplete, and subject to dynamic changes. Methods to manually structure web content based on user-defined attributes have been proposed in the past \cite{dontcheva_summarizing_2006}. For SB-users, it may be desirable though, to have a more automated approach to perform this kind of drill-down.

\subsection{Personal Information Collection}

When using IC over a longer period of time, some users gradually built a personal information collection of potentially interesting resources covering multiple topics. These topics can be organized hierarchically, as shown in Figure \ref{fig:keywords_subtopics}. Manual spatial organization of the gathered documents seems to be an intuitive and appropriate method to express this hierarchical organization.

\begin{figure}[h]
 \centering
   \includegraphics[width=0.5\textwidth]{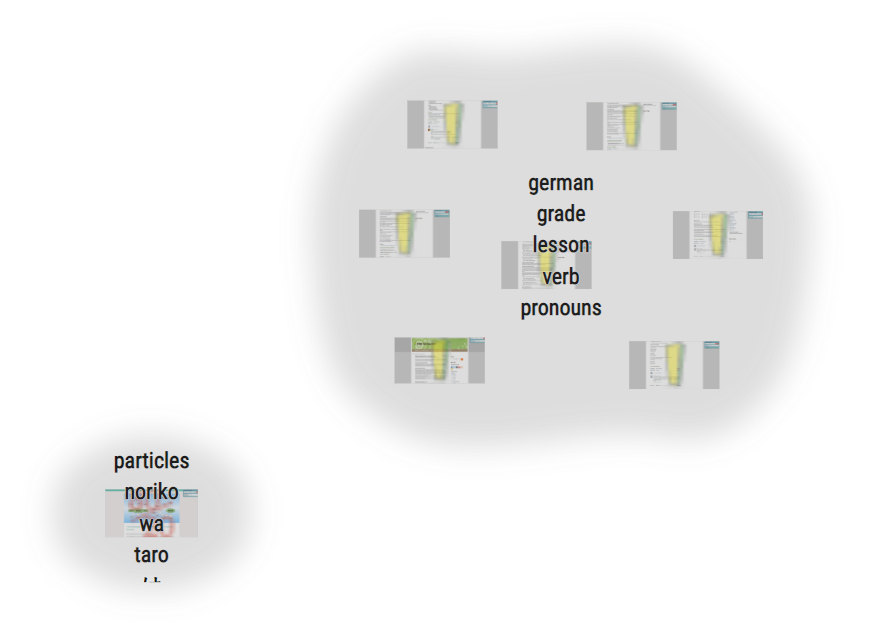}
  \caption{Zooming into the cluster containing documents about the Japanese language by user LN in Figure \ref{fig:bookmarks} (bottom right) reveals two sub-topics.  }
  \label{fig:keywords_subtopics}
\end{figure}

To reveal the hierarchical topic structure, semantic zooming was received as valuable feature. However, the simple tf*idf method we used to extract keywords to label clusters of information fragments in IC was not expressive enough, especially for complex and whole-document snippets. For more expressive content summaries, titles of the captured web sites, their domain names, and -- if available -- the query terms used to find the web site could be taken into account additionally. The web sites' content structure could furthermore indicate importance of keywords, so that keywords contained in headers are higher prioritized. Further experimentation is required to explore how users would manually summarize information fragment content, to which extent these manual labels can be automatically reconstructed from the different sources of text information, and how much user intervention is required to reach the expected level of expressiveness.

\section{Conclusions}

Information- and knowledge-intensive work requires a lot of organization and structuring of information to come from multiple raw information sources to a product, such as an article or design guidelines. We introduced the concept of an ``information collage'', where users can freely organize information fragments extracted from arbitrary information sources in a zoomable 2D space. Results from our evaluation show that such a concept is particularly useful to create a persistent personal information collection covering multiple, hierarchically organized topics. Our evaluation also showed that users perform information collection and structuring in fundamentally different ways, and each usage strategy requires different interaction and visualization techniques. To create an information and knowledge map, users need powerful methods for manual information structuring. To create and explore a ``shoebox'' of relevant documents, users expect automatic document categorizations and summaries, as well as powerful interaction techniques to extract relevant material from the documents in the shoebox. As the shoebox usage strategy was most commonly observed in our study, it seems promising to explore a suitable balance between automatic content-based categorization of stored information and manual information structuring in the future. 

\section*{Acknowledgements}

This work was financed by the Austrian Science Fund (FWF): T 752-N30. 
We thank all users of our studies for their support.

\end{techReport}

\printbibliography[heading=bibintoc]

\end{document}